\documentstyle[12pt,aasms]{article}
\slugcomment{submitted to {\sl ApJ}}

\newcommand {\eg} {e.g.}		% e.g.
\newcommand {\etal} {et al.}		% et al.

\begin{document}

\title{Age Gradient and the Second Parameter Problem in the Galactic Halo}

\author{Harvey B. Richer\altaffilmark{1}, William E. Harris\altaffilmark{2}, Gregory G. Fahlman\altaffilmark{1}, Roger A. Bell\altaffilmark{3}, Howard E. Bond\altaffilmark{4}, James E. Hesser\altaffilmark{5}, Steve Holland\altaffilmark{1}, Carlton Pryor\altaffilmark{6}, Peter B. Stetson\altaffilmark{5}, Don A. VandenBerg\altaffilmark{7} \& Sidney van den Bergh\altaffilmark{5}}

\altaffiltext{1}{Department of Geophysics \& Astronomy, 
University of British Columbia, 
Vancouver, B.C., 
V6T 1Z4.  E-mail
surname@astro.ubc.ca}

\altaffiltext{2}{McMaster University, Department of Physics and Astronomy, Hamilton,
ON, Canada L8S 4M1. E-mail harris@physun.physics.mcmaster.ca}

\altaffiltext{3}{University of Maryland, Department of Astronomy, College Park, MD
20742--2421. E-mail rabell@astro.umd.edu}

\altaffiltext{4}{Space Telescope Science Institute, 3700 San Martin Drive, Baltimore,
MD 21218. E-mail bond@stsci.edu}

\altaffiltext{5}{Dominion Astrophysical Observatory, Herzberg Institute of
Astrophysics, National Research Council, 5071 W. Saanich Road, RR5,
Victoria, B.C., Canada V8X 4M6. E-mail surname@dao.nrc.ca}

\altaffiltext{6}{Rutgers University, Department of Physics and Astronomy, PO Box 849,
Piscataway, NJ 08855--5462. E-mail pryor@physics.rutgers.edu}

\altaffiltext{7}{University of Victoria, Department of Physics and Astronomy, PO Box
3055, Victoria, BC, Canada V8W 3P7. E-mail davb@uvvm.uvic.ca}

\begin{abstract}

We establish a framework for determing absolute ages of Galactic
globular clusters and then use these ages to investigate the
age-metallicity and age-Galactocentric distance relations for the 36
clusters with the most reliable age data. The clusters span
Galactocentric distances from 4 through 100 kpc and cover a metallicity
range from $[Fe/H] = -0.6$ to $-2.3$.  Adopting currently plausible choices for
the relation between cluster metallicity and horizontal-branch luminosity,
and alpha-enhancement ratios, we find that the majority of the globular
clusters form an age distribution with a dispersion $\sigma(t)$
about $10^9$ years, and a total age spread smaller than 4 Gyr. Clusters in the
lowest metallicity group ($[Fe/H] < -1.8$) appear to be the same age to well
within 1 Gyr at all locations in the Milky Way halo, suggesting that star
formation began throughout the halo nearly simultaneously in its earliest 
stages. We find no statistically significant correlation between mean
cluster age and Galactocentric distance (no age gradient) from 4 to 100
kpc.   
The correlation between cluster
age and horizontal-branch type suggests that causes in addition to
metallicity and age are required to understand the distribution of
stars along the horizontal branches in globular cluster
color-magnitude diagrams.

\end{abstract}


%%%%%%%%%%%%%%%%%%%%%%%%%%%%%%%%%%%%%%%%%%%%%%%%%%%%%%%%%%%%%%%%%%%%%%%%

\section{Introduction}

 It is not yet clear how, and in which order, the oldest constituents
of the      Galaxy formed.  The current debate is bounded by two well
known extreme      scenarios:  one is the ELS rapid-collapse model
(Eggen, Lynden-Bell \& Sandage      1962; Sandage 1990) in which the bulk
of star formation in the halo occurred over      not much more than a
rotation period.  In the simplest form of the ELS picture, 
all the globular clusters      might 
have a rather small range in ages, and cluster age and     
metallicity are not expected to be strongly correlated with
Galactocentric distance.      Alternatively, Searle \& Zinn (1978 = SZ) 
proposed that both the      halo clusters and the halo
field stars formed in fragments (possibly originally      located
outside the Milky Way) which had their own individual histories of star 
    formation and chemical enrichment.  In addition, another possible
consequence of this picture is that the accretion of major fragments 
could have continued for several Gyr following the initial collapse.

The original motivation
for the SZ picture was to explain in a natural way the
wide range of globular cluster metallicities that we observe at all
locations in the halo, as well as the progressive emergence of
the enigmatic  `second parameter' (see below)
with increasing Galactocentric distance.
Thus, in the SZ scenario, the halo
cluster system might exhibit a significant age      spread, especially in
its outermost regions where the various 
second-parameter anomalies are strongest.  
In this connection, Lin \& Richer (1992) 
have noted that some surprisingly      young
outer-halo clusters could have been captured more recently from
satellites      of the Milky Way, rather than formed within the
environs of the Milky Way halo. This view has been
reinforced by the recent observations of the 
Sagittarius dwarf and its clusters (Ibata, Gilmore \& Irwin 1994; 
Da Costa \& Armandroff 1995), which appear to be actively undergoing
accretion into the Milky Way halo at the present time. 
Therefore, the existence of a few such
young clusters need not necessarily be a signature of an extended phase
of star formation within the Galactic halo itself.  On the theoretical
side, a quantitative
model of globular cluster formation has been developed by Harris \& Pudritz
(1994) and McLaughlin \& Pudritz (1996) which identifies the SZ gaseous
`fragments' as supergiant molecular clouds (SGMC):  
essentially $10^8 - 10^9 M_{\odot}$ 
versions of the smaller GMCs that reside in the Galactic disk today.
A specific consequence of this model (in which protoclusters 
are postulated to build up by
the collisional accretion of small cloudlets within the host GMC)
is that the growth time should increase outward in the halo, where the
ambient gas pressure and density are smaller. Thus, we 
should expect to see a larger range of cluster ages in the outermost halo
and (conversely) a very small range (less than 1 Gyr; see McLaughlin \& Pudritz
for quantitative predictions)
in the inner bulge where the collisional growth times
are fastest.  When this picture is added to the possibility of 
late infall and accretion as mentioned above,
we might expect the total age distribution of the globular 
clusters in a large galaxy such as the Milky Way 
to be a complex story indeed.

          On the observational side, the
`second parameter' problem in the color-magnitude diagrams
of globular clusters remains a keystone to understanding the cluster age
distribution. Among globular cluster horizontal branches (HB), the general trend is for them to be redder than the RR Lyrae instability strip in clusters with $[Fe/H] > -1$, and for the color distribution to become increasingly bluer with decreasing $[Fe/H]$. Almost all clusters with $[Fe/H] < -2$ have HBs dominated by stars bluer than the instability strip. There are, however, exceptions to this rule with a number of clusters possessing HB morphologies too red or too blue for their $[Fe/H]$ values. Hence there must be some additional parameter(s) which influence HB morphology. This is the `second-parameter' problem. Thorough recent reviews of this problem  
are given by Carney, Fullton \& Trammell
(1991), van den Bergh (1993), Lee, Demarque \& Zinn (1994 = LDZ) and Chaboyer, Demarque \& Sarajedini (1995 = CDS).   
The current standard solution of this problem is cluster age. With all other parameters held constant, the HB of a cluster becomes bluer with increasing age due to decreasing envelope mass. Observationally it is known that the second-parameter effect becomes more evident at large Galactocentric distances $R_{gc}$, where many clusters possess HBs that are redder than the mean for their $[Fe/H]$ values. Hence if the second parameter is indeed age, the average age of the outer halo clusters should be smaller than that of the nearby ones, and the outer halo clusters should simultaneously exhibit a larger age spread. It is important to stress that this is one of the main underpinnings of the SZ scenario for the chaotic formation of the halo and for Zinn's (1993) recent revision of this model.
 
Evidence in favor of the view that age is the second parameter has been developed through
the detailed simulations of cluster HBs by LDZ. They argue that the age of the halo must decrease monotonically from the center of the Milky Way outward. 
Other     parameters, such as the CNO-group abundances, helium abundance,
mass loss in pre-HB stages, core rotation, or possibly even cluster
density (Fusi-Pecci \etal\ 1993; but see van den Bergh \& Morris
1993), may all have effects on the temperature      distribution of stars
along the HB.  However, the model simulations of 
LDZ show that most of these others 
are unlikely to be the `global' second parameter, with age remaining as the
main contender.   

Unfortunately, these conclusions remain at least partly 
circumstantial since the morphologies of the HB and red-giant branches
by themselves do not directly measure cluster age.
The most direct approach to establishing cluster ages is to 
obtain deep color-magnitude photometry of the main-sequence stars.  
Although many data of this type are available for the nearby 
globular clusters (see, e.g. VandenBerg, Bolte \& Stetson 1990 = VBS),
sufficiently accurate photometry of the turnoff and unevolved main sequence
for the most remote halo clusters (and thus the most extreme second-parameter
anomalies) has proved beyond the reach of current ground-based telescopes.

Recently, Stetson \etal\ (1995) have used the Hubble
Space Telescope to obtain highly accurate deep CMDs for three of
the outermost halo clusters, representing the extremes of the metallicity
range found there.  The purpose of our paper is to show that
the new results of Stetson \etal, 
together with the most reliable age calibrations for 
globular clusters elsewhere in the halo,  
indicate that there exist very distant halo clusters that are 
as old as clusters in the inner halo. 
Furthermore, we show that, with entirely plausible assumptions concerning
the cluster abundances and distance scale, there is no 
net age gradient in the Galactic halo,
and the cluster-to-cluster age dispersion, particularly for clusters at the
same metallicity, is remarkably small. 
The notable exceptions to these results comprise a handful of strongly
anomalous clusters at intermediate $R_{gc}$ which may be late-accretion
objects and which form a diverse lot even among themselves (see \S3.3 below).
In its entirety, this view is different from that suggested by LDZ but similar to CDS, and
it remains to be seen how it can be accommodated 
in detail by Galactic formation models.

%%%%%%%%%%%%%%%%%%%%%%%%%%%%%%%%%%%%%%%%%%%%%%%%%%%%%%%%%%%%%%%%%%%%%%%%%

\section{Data}

VBS have examined the best     
existing CMDs for globular clusters in considerable detail 
and have also determined a homogeneous set of relative
ages for clusters within several      metallicity sub-groups.  VBS adopt
the position that, due to uncertainties in the relevant physics (\eg,
detailed chemical abundance ratios, convection, opacities), it is
dangerous to use theoretical models to compare absolute ages across
different metallicity regimes by their color-based technique. On the
other      hand, relative ages can be determined differentially with
high precision (as small as 0.5 Gyr for the best-studied objects) 
with respect to fiducial clusters
within narrowly defined metallicity sub-groups.  

To form a list of clusters with the most accurately estimated ages, we
begin      with objects in the VBS compilation.  To these we add several
other clusters with      more recently published high quality, 
main-sequence photometry:  NGC 7078      (Durrell \& Harris 1993), NGC
6101 (Sarajedini \& da Costa 1991), NGC 5053      (Fahlman, Richer \&
Nemec 1991), NGC 1904 (Chaboyer, Sarajedini \& Demarque      1992),
Ruprecht 106 (Buonanno \etal\ 1990), Arp 2 (Buonanno \etal\ 1994), NGC  
   1851 (Walker 1992b), NGC 6229 (Buonanno 1994), NGC 6352 (Fullton \etal\ 1995),
and our recent HST study of NGC 2419, Pal 3  and Pal 4
(Stetson \etal\ 1995). For
NGC 6352 the differential age
measurement is made from the difference in magnitude between the
horizontal branch and the cluster turnoff      (Buonanno \etal\ 1989), 
rather than from the VBS color-differential technique. 

Our primary goal is to use this material to investigate the 
correlations between cluster age and several other parameters
including metallicity, HB type, and (perhaps most interesting)
Galactocentric distance.  The key distinction between our results
and those of (for example) LDZ is that the ages are deduced directly from
the main sequences, rather than indirectly from HB and RGB morphology.
However, to intercompare clusters with very different metallicities
and convert our differential ages into absolute ones, we must make
some further assumption about the age zero points in each metallicity
group.  To set the scale for absolute ages, we extract from the
literature the      age estimates derived from full-scale
isochrone fitting for selected clusters in each metallicity      group,
and then use the
differential (VBS) results to deduce ages for the      remaining
clusters in that metallicity group.  To ensure that the
data are homogeneous (i.e., similar input physics for the isochrones), 
we have, wherever possible, used clusters whose absolute
ages were      determined from detailed isochrone fits with the
Bergbusch-VandenBerg (1992)      oxygen-enhanced isochrones and with
helium abundances in the range $Y = 0.23 -      0.25$.  Although
we do not use the HB to establish either the distance or age, 
it turns out that the distance scale
for our selected fiducial clusters (see below) is essentially     
equivalent to an HB luminosity calibration,
\begin{equation}
M_{V}(RR) = M_{V}(HB) = 0.15 [Fe/H] + 1.0
\end{equation}
\noindent 
which is in accord with the great majority of current
evidence (see LDZ; CDS; Carney, Storm \& Jones 1992 = CSJ;
Skillen \etal\      1993).
(Note, however, that RR Lyrae observations in the LMC (Walker, 1992a), in
conjunction with Cepheid distances to the Large Cloud, yield
significantly brighter HB luminosities for the old component of the
LMC; if a similarly bright level were to hold for the outer globular
clusters in the Milky Way, their ages would be correspondingly 
reduced.)

          Following VBS, we employ four metallicity subgroups, for 
which the age zero points      are set as follows.

     (1)  $[Fe/H] <  -1.8$

          Fiducial clusters in this especially well defined group are
NGC 4590      (McClure \etal\ 1987), NGC 6341 (Stetson \& Harris 1988) 
and NGC 7099 (Bolte 1987a).  The
{\it relative} age determinations for all      three of these clusters (see VBS) imply that they have      the same age to
within 0.5 Gyr.  However, even though the papers listed all used O-enhanced isochrones     
($[O/Fe] = 0.7$) and have similar precision, the derived absolute ages range from 14 to 17 Gyrs. This illustrates clearly the difficulty in establishing absolute ages for globular clusters.  Averaging the results, we adopt an
absolute      age of 16 ($\pm 1.6$) Gyr for NGC 7099, from which we derive the
values for all of the      other clusters in the group differentially.

     (2)  $-1.8 < [Fe/H] < -1.5$

          Our single fiducial cluster is NGC 7492, with an age of 15 (${\pm 2}$)
Gyrs from the      same O-enhanced isochrones ({C\^ot\'e} \etal\ 1991).

     (3)  $-1.5 < [Fe/H] < -1.00$

          In this group, there are, unfortunately, no totally reliable
CMDs that have      been well fitted to the O-enhanced isochrones.  We
therefore chose to go a bit      further back in the literature and
selected age determinations from the VandenBerg      and Bell (1985)
scaled-solar isochrones.  These are available in a consistent way     
for NGC 362 (Bolte 1987b) and NGC 5904     
(Richer \& Fahlman 1987).  From each
of      these, we subtract 2 Gyr to bring them back to the equivalent
age      for $[O/Fe] = +0.7$ as used in the previous two groups (see,  \eg,
Durrell \& Harris      1993).  This procedure yields an average absolute age of 
14 (${\pm 1.4}$) Gyr for these two clusters.      NGC 288, with its fainter turnoff point and extremely blue
HB, is also in this      group, and is about 2
Gyr older, according to the differential age calculation      (VBS).

     (4)  $-1.0 < [Fe/H]$

          Absolute age determinations for 47 Tuc (Hesser \etal\ 1987)
and NGC      6838 (Hodder \etal\ 1992) give a mean age of 14 ($\pm 1$) Gyr for
these two metal-rich      objects.  The isochrones here use $[O/Fe] = 0.3
- 0.4$, in accord with the available      data for halo field star
abundances.

          Data for a total of 36 globular clusters with Galactocentric
distances      ranging from 4 to 100 kpc are included in the sample and
are listed in Table 1.      Unfortunately, no clusters within 3 kpc of
the Galactic nucleus yet have CMDs that are good
enough to allow accurate ages to be determined. Metallicities,
distances, and HB morphology indices  are taken from the 
recent compilation of Harris (1995).

The pitfalls of estimating absolute ages for clusters are well known and
have been emphasized in the literature many times (see VBS or Bolte \& Hogan 1995 for 
 particularly extensive discussions).  
Even small uncertainties in the adopted cluster
reddening, distance, or      composition can lead to uncertainties in the
deduced absolute age that are typically 2 Gyr for clusters with well
defined main-sequence photometry, such as those included here.  
By averaging over as many clusters as possible, and by using a      uniform
distance scale with the same set of isochrones, we may then reasonably
attempt to reduce the uncertainty in the zero point for each subgroup
(relative to the other subgroups) to $\simeq 1 - 2$ Gyr.
The differential age measurements within a subgroup have internal
uncertainties (precisions) of typically 0.5 Gyr for the clusters
included here, all of which have high-quality photometry  
(see VBS, Durrell \& Harris, and Stetson \etal).  The
discussion in the following sections demonstrates that this expectation is
achievable.

%%%%%%%%%%%%%%%%%%%%%%%%%%%%%%%%%%%%%%%%%%%%%%%%%%%%%%%%%%%%%%%%%%%%%%%%%

\section{Ages, Metallicities and the Second Parameter}

The discussion of LDZ, which establishes relative cluster ages by
careful interpretation of HB morphology, 
suggests that there is a net
age gradient through the Milky Way halo amounting to 
2 Gyr over 40 kpc and 4 Gyr over 100      kpc.
We wish to investigate this same question, relying instead on ages
determined from the main-sequence turnoff region of the CMD.
However, two questions we must address first are:
(1) is there an age-metallicity relation
for      globular clusters, and (2) to what extent does 
(turnoff-calibrated) age correlate with HB morphology?

\subsection{The Age-Metallicity Relation}

The age-metallicity correlation for the present data is shown in Fig. 1. Since the data points at a given metallicity are not independent, as they were determined differentially with respect to one of the clusters, we took an average of the
ages and metallicities in each bin using the same weights for the
metallicities as for the ages. This resulted in four uncorrelated points, one for each
metallicity bin, with the $\pm 1\sigma$ error bars in the ages corresponding to two sources of error. These are the
uncertainty of the absolute zero-point for each bin and the uncertainty in
the differential age for each cluster (estimated from the uncertainty
in measuring the shift in the position of the red-giant branch from the
reference cluster to the cluster in question which was taken
to be 0.5 Gyr for all the globular clusters in our sample). The errors from these sources were
then added as follows to produce a final error in each of the four points, 
\begin{equation}
\sigma^2(total) = \sigma^2(absolute) + \sigma^2(differential)/N.
\end{equation}
\noindent
where N is the number of clusters in a metallicity bin.  
These points (filled circles), together with their error bars are plotted on
top of the individual cluster points (open circles) in Fig. 1.  These four data points  should be
effectively uncorrelated, so simple propagation of error should do to
estimate the error in the slope.

Excluding the four `young' clusters with ages $< 12$ Gyr 
that may have been accreted from other systems (see 
\S3.3), we
find a slope in the Age-[Fe/H] plane which is different from zero, but it is significant at less than the $1\sigma$ level (68\% confidence). Formally, a linear least squares fit to the data, excluding the young systems, yields 
\begin{equation}
Age(Gyr) = -1.21(\pm 1.33)[Fe/H] + 13.04(\pm 1.74).
\end{equation}
\noindent
This slope should be compared with that of $-4.0$ derived by CDS. To check our result, we carried out Monte-Carlo simulations of the data allowing the individual cluster differential ages to vary randomly by $\pm 0.5$ Gyr and the absolute age to vary by $\pm 1\sigma$ where $\sigma$ is the absolute age error for the metallicity group. We then reconstructed the four data points and carried out least squares fits to them. A total of $10^4$ trials were done in this manner and they yielded a mean slope of $-1.22(\pm 1.35)$, completely consistent with our initial results.  

The 11 clusters with $[Fe/H] <  -1.8$ have a     
standard deviation in age of only 0.4 Gyr, which is to be compared to an age dispersion
of 0.9 Gyr for the 21 clusters with $[Fe/H] > -1.8$.   
Taken at face value, our data are then
consistent with a      scenario in which the
most metal-poor clusters formed initially, over a length of time consistent with the
rotation period of the Galaxy. The more metal-rich clusters formed somewhat later over a more extended period of time. The confidence in such a scenario is low, however, due to the low statistical significance of the slope in the Age-[Fe/H] relation. Our data are not strongly inconsistent with a picture in which all clusters of all metallicities formed simultaneously.

However, the results in the preceeding paragraph  are  the
least secure of the several presented in this paper because they
depend on the variation of oxygen and the other
alpha-elements with $[Fe/H]$.  The Bergbusch \& VandenBerg (1992)
calculations do not, in fact, allow for the observed enhancements in
metal-poor stars ({\it cf}. Wheeler, Sneden, \& Truran 1988) of any of the
alpha-elements except oxygen (for reasons mentioned in their paper). 
Moreover, their adopted $[O/Fe]$ values now appear to be somewhat too high
when compared with the latest determinations of this quantity in field
subdwarfs and subgiants.  For instance, whereas Bergbusch and VandenBerg
adopted $[O/Fe] = 0.7$ at $[Fe/H]  = -2$, current indications are that field
stars with this iron content have $[O/Fe]$ somewhere in the range of 0.4
to 0.6 (e.g., Spite \& Spite 1991; Bessell, Sutherland \& Ruan 1991). 
However, the high $[O/Fe]$ values assumed by Bergbusch \& VandenBerg
compensate to some extent for their neglect of enhancements in the
abundances of other alpha-elements.  New computations by VandenBerg  \etal\ 1995, which use the latest Livermore opacities, indicate that (for
$[Fe/H] = -2$) the turnoff luminosity versus age relations for $[O/Fe] =
0.7$, with scaled solar number abundance ratios for all other elements,
correspond closely to those for [alpha$/Fe] = 0.5$, where `alpha' = O,
Ne, Mg, Si, etc.  Consequently, there is some basis for confidence in
the ages which we have inferred for the globulars using the Bergbusch
and VandenBerg isochrones.

Among recent similar discussions of the age-metallicity relation for
globular clusters, the most nearly comparable ones to ours are probably
that of CSJ (compare their Figs. 20-21 with our Fig. 1) and CDS (compare their Fig. 1 with our Fig. 1). 
Aside from a small zero-point difference in the age scale due to their
particular choices of [alpha$/Fe]$ and the isochrones, CSJ find
substantially similar results compared with ours -- no clear trend of age
with metallicity for $[Fe/H] > -1.8$, and a higher age dispersion for the
intermediate-metallicty group. Notably, however, both CSJ and CDS also find
significantly larger ages than we do for the most metal-poor group
($[Fe/H] < -1.8$) and thus conclude that a clear overall age-metallicity
relation exists.  We do find such a relation, but it is certainly not a very robust result, as shown in Fig. 1. CSJ find that
an increased $[O/Fe]$ ratio for these metal-poorest
clusters would reduce, though not eliminate, the relation, leaving the
metal-poor group about 2 Gyr older than the mean of the other clusters. CDS, however, claim to detect a very significant age-metallicity relation for the clusters with the most metal-poor systems again being the oldest.
The difference between these results and ours appears to be due to a
combination of small effects, since our approach differs in several
details.  Both CSJ and CDS first adopt a distance to
each cluster from the HB level. From
this CSJ calculate the bolometric magnitude of the turnoff point, then
finally the cluster age from $M_{bol}$(TO) and the adopted stellar models. Similarly, CDS derive the cluster age from model calibrations of $M_V$(TO)  as a function of metallicity, where $M_V$(TO) is determined from the difference between the level of the HB and the TO, $\Delta V_{TO}^{HB}$.
By contrast, the ages we use are deduced from full best-fit isochrone
comparisons to the main-sequence, turnoff, and subgiant regions of the
fiducial clusters in each metallicity group, followed by differential
age determinations of the other clusters in the group with the
method of VBS.  Although our results are, within the observational
errors, {\it consistent} with the CSJ and CDS distance scales (equation
1), we do not use the HB level to set either the distance or the age.

Discussions by Sandage \& Cacciari 1990 and Sandage 1993 demonstrate how
the age-metallicity relation
changes as a function of the adopted $M_V(HB)$ distance scale,
when the HB (or RR Lyrae) stars are used to
set the cluster distance. The turnoff luminosity, and finally
the cluster age are derived through the model dependence of $M_{bol}$(TO) on age.
Although they favor a notably steeper slope on $M_V$(HB) vs. $[Fe/H]$
(see CSJ for an exhaustive discussion of the HB luminosity
calibrations), they find mean age differences of $\sim 2$ Gyr across
the full metallicity range.  Interestingly, the closest analog of their
study to ours is through their alternate derivation of the cluster
distances by direct main-sequence fitting to the stellar models,
from which they find no trend of age with metallicity and a mean age
of 15.5 Gyr with $[O/Fe] = 0.6$ (see Fig. 14 of Sandage \& Cacciari).

\subsection{The HB Type - Age Relation}

As another way of exhibiting the age dispersion and mean age of our
cluster sample, we show in Fig. 2 a plot of cluster age 
versus population gradient along the HB.  Let us first
concentrate on the central part of this distribution, excluding the
clusters with extreme blue or red HBs.  The nine objects with
$-0.8 < (B-R)/(B+V+R) < +0.4$ are
seen to have an rms age dispersion of just 0.6 Gyr. (B, V, and R are defined as the number of HB stars to the blue (B), red (R) and in the variable (V) region of the HB of the cluster. More details can be found in LDZ.) 
Such a cluster-to-cluster dispersion is the smallest it
could be when the differential age measurement 
uncertainties are taken into account. 

A complementary way to view the distribution in Fig. 2 is that
seven of these same nine clusters
have metallicities in the narrow range $-1.45 < [Fe/H] < -1.26$, have an extremely small dispersion in age of 0.48 Gyr
and yet span 60\% of the entire range in HB type. From the LDZ simulations,
at $[Fe/H] = -1.2$ the rate of change of the HB parameter with age is $\Delta(HB)$/$\Delta(t) = 0.3$/Gyr while at an age of 13 Gyr $\Delta(HB)$/$\Delta([Fe/H]) = 2.46$/dex. The observed dispersion in the HB parameter for these seven clusters is 0.41, the age dispersion is 0.48 Gyr while the dispersion in [Fe/H] is a mere 0.07. Hence, it appears that the dispersion in the HB parameter caused solely by age (0.14), caused solely by the metallicity dispersion (0.17) or due to both effects (0.22) is too small to explain the observed spread. 
This suggests {\it that variations in HB morphology for these
objects are not primarily due to differences in age and
metallicity}.  

It is only at the extremes of the distribution in HB type
that a larger dispersion in age is seen. Red HB clusters span an $\sim 5$ 
Gyr range in ages, although considering the objects at 
$(B-R)/(B+V+R) \sim -1$ as a single group is probably ill-advised,
since it includes both the normal metal-rich bulge clusters, 
such as NGC 6352 and M71, and outer-halo objects, 
such as Rup 106 and Pal 4, which must be generically very different
(see below). The blue-HB clusters for which $(B-R)/(B+V+R) \sim +1$ 
tend to have somewhat older ages, with the important exception
that Arp 2, a pure blue HB cluster, appears to be very young.  These results
clearly indicate that, even at a single metallicity, there is no
one-to-one correspondence between the ages of globular clusters and
their HB types which can be applied globally.

\subsection{Is There An Age Gradient in the Galactic Halo?}

To examine evidence for or against an age gradient in the Galactic
halo, we      plot in Fig. 3 the derived absolute ages versus
Galactocentric distances for the      clusters listed in Table 1.  
Two features of this diagram immediately stand out:  (a) The great
majority of the clusters, over all metallicity groups, fall within a
single band with a mean age at 14.9 Gyr, an rms dispersion of 1.2 Gyr,
and a total range of 3.7 Gyr.  (b) Four clusters 
(Pal 12, Rup 106, Arp 2, and Ter 7) stand distinctly off this
main band, with ages that are younger by $\sim 5$ Gyr than the main group.
In addition, the latter handful of objects is already somewhat 
over-represented, 
because a much larger fraction of the outer halo clusters have now been investigated than the inner halo ones. Thus these `young' objects can arguably be
viewed as having anomalously low ages.

We comment first on the age dispersions. Within the main cluster
population as defined by Fig. 3, the overall dispersion is already
small, but it is strikingly lower 
among clusters within the same metallicity group.
For      the most metal-poor group, the dispersion is 0.4
Gyr, while in the      worst case (for the clusters at $[Fe/H] \sim -1.2$),
the dispersion is still only 0.9 Gyr.      This larger dispersion is due
mainly to NGC 288 and NGC 6254, which appear to      be 2.0 and 1.5 Gyr
older than the mean of the other clusters at this metal      abundance.
As implied in the preceding sections, this narrow age spread is the
direct result of our particular approach to 
fitting the adopted isochrones, along with the chemical composition
parameters.  However, the main point we wish to stress is that,
using a highly plausible set of assumptions and model fitting
methodology along with the best available cluster data, we find that it
is possible to argue that the cluster-to-cluster differences in age, particularly at a given metallicity, 
are as small as they could possibly be expected to be, given the 
internal precisions (0.5 Gyr at least) in the age determinations
themselves.

Next we address the question of an overall age gradient.
Using the entire sample of clusters with $R_{gc}$ from 4 to 100 kpc (and excluding the clearly anomalous young
objects, which are discussed below)
there is no evidence for a radial age 
gradient.  A linear least-squares fit to the data for these clusters yields
a slope of $-0.001 \pm 0.011$ Gyr/kpc.  The three outermost clusters in
our sample (NGC 2419, Pal 3, Pal 4) have extended the data to $R_{gc} =
100$ kpc, whereas the CDS compilation  contained no systems beyond 40 kpc. The existence of these clusters in the far outer halo of the Galaxy at ages identical to those in the inner regions, demonstrates that star
formation, even in this remotest part of the Galactic halo, began just as
early as in the considerably denser inner halo.
The outermost halo remains a region of considerable interest,
and there remain several 
globular clusters beyond 30 kpc      from the Galactic center for which
no accurate ages are yet available.  We are currently 
obtaining new age determinations      for some of these other
very remote clusters.

Inspection of Fig. 3 also suggests that the main era of globular cluster
formation may have ended rather abruptly   $\sim 13.3$ Gyr ago.  The
actual numerical      value for the time at which cluster formation
ended of course depends on the zero      point from the particular set
of isochrones and abundance ratios that we have      adopted.  The
important feature is the surprisingly sharp lower edge to the     
distribution.  It is highly unlikely that this feature can 
have been generated accidentally or as a byproduct of 
large random errors in the age determinations, which
would be expected to blur out a true physical cutoff.  This provides
additional evidence that the differential age determinations are indeed
at the claimed level of 0.5 - 1 Gyr for clusters with high-quality
photometry.

In the preceding discussion, the four anomalously young clusters with
intermediate Galactocentric distances ($R_{gc} \sim 20$ kpc)
have been neglected.  These 
clusters share      several properties which appear to set them apart
from both the main group of clusters in Fig. 3 and the outermost-halo
ones, such as Pal 4.   (1)  Buonanno et al. (1994) 
and Da Costa \& Armandroff (1995) argue that three of these clusters,
along with NGC 6715, are associated with the
Sagittarius dwarf galaxy (Ibata \etal\ 1994). 
This suggests that most of these small, younger clusters
may      have been captured long after
their formation elsewhere. (2)  Possibly confirming evidence of this is that they have much
lower luminosities than the typical cluster in the Galaxy; their mean $M_V$ is $-5.8$ (Webbink      1985), whereas
it is $-7.4$ for the globular cluster system of the Galaxy as a whole.    
As Harris \& Pudritz (1994) note, they might therefore have formed
within parent      SGMCs that were an order of magnitude smaller than
those making up the original  proto-halo. (3)  Further support for their formation elsewhere is that more than
65\% of the      globular clusters in the Galaxy have core radii less
than 2 pc, whereas all the `young'      clusters have larger cores, with
Arp 2 possessing one of the largest known cores at      16 pc radius
(Webbink 1985).  

These bits of evidence, though      clearly
circumstantial, suggest to us that these four clusters should not be     
considered when discussing the early formation history of the Milky Way.
What      they certainly do indicate is that the history of the
Galactic halo is an ongoing      process with 
satellite galaxies occasionally being absorbed into the Milky Way up to the present time.

%%%%%%%%%%%%%%%%%%%%%%%%%%%%%%%%%%%%%%%%%%%%%%%%%%%%%%%%%%%%%%%%%%%%%%%%%%%%%%

\section{Summary and Conclusions}

The absolute age distribution of globular clusters points to a number
of      conclusions which are likely to bear on the manner in which the
Milky Way Galaxy      formed. 

 (1)  The most metal-poor clusters may be slightly older than clusters at higher metallicities. These metal poor systems formed on a short timescale that was  similar to the free-fall collapse time of the proto-Galaxy.  It appears that these clusters formed throughout the entire halo of the Galaxy at very nearly the same time. 
  
(2) There is no evidence for a gradient in globular cluster ages for those systems lying between about 4 and 100 kpc
from the Galactic center. It thus appears that when clusters formed in the Galaxy, they did so throughout its entire extent. This is quite a remarkable result implying that the physical conditions capable of supporting cluster formation existed over a huge distance and hence presumably over a wide range in physical properties.  

(3)  The HB morphology of clusters must, at least in part,  
be due to causes other than metallicity and age.  Furthermore,
interpreting the      systematic trend toward redder HB morphology at
increased $R_{gc}$ as being primarily due to age (LDZ) appears to be
too simplistic. 

Finally, we note, that accurate ages for the innermost (bulge-type, high-
     metallicity) clusters are not yet available in
large numbers, and may still change
the overall      conclusions once they can be added to the correlations
that we have discussed      above.

%%%%%%%%%%%%%%%%%%%%%%%%%%%%%%%%%%%%%%%%%%%%%%%%%%%%%%%%%%%%%%%%%%%%%%%%%%%%%%

\acknowledgments

The research of HBR, WEH, GGF, and DAV is supported through grants from
the Natural Sciences and Engineering Research Council of Canada.

%%%%%%%%%%%%%%%%%%%%%%%%%%%%%%%%%%%%%%%%%%%%%%%%%%%%%%%%%%%%%%%%%%%%%%%%%%%%%%

\clearpage

%%%%%%%%%%%%%%%%%%%%%%%%%%%%%%%%%%%%%%%%%%%%%%%%%%%%%%%%%%%%%%%%%%%%%%%%%%%%%%%

%\null
%\bigskip
%\centerline{\bf FIGURE CAPTIONS}
%\medskip

\begin{figure}
\plotone{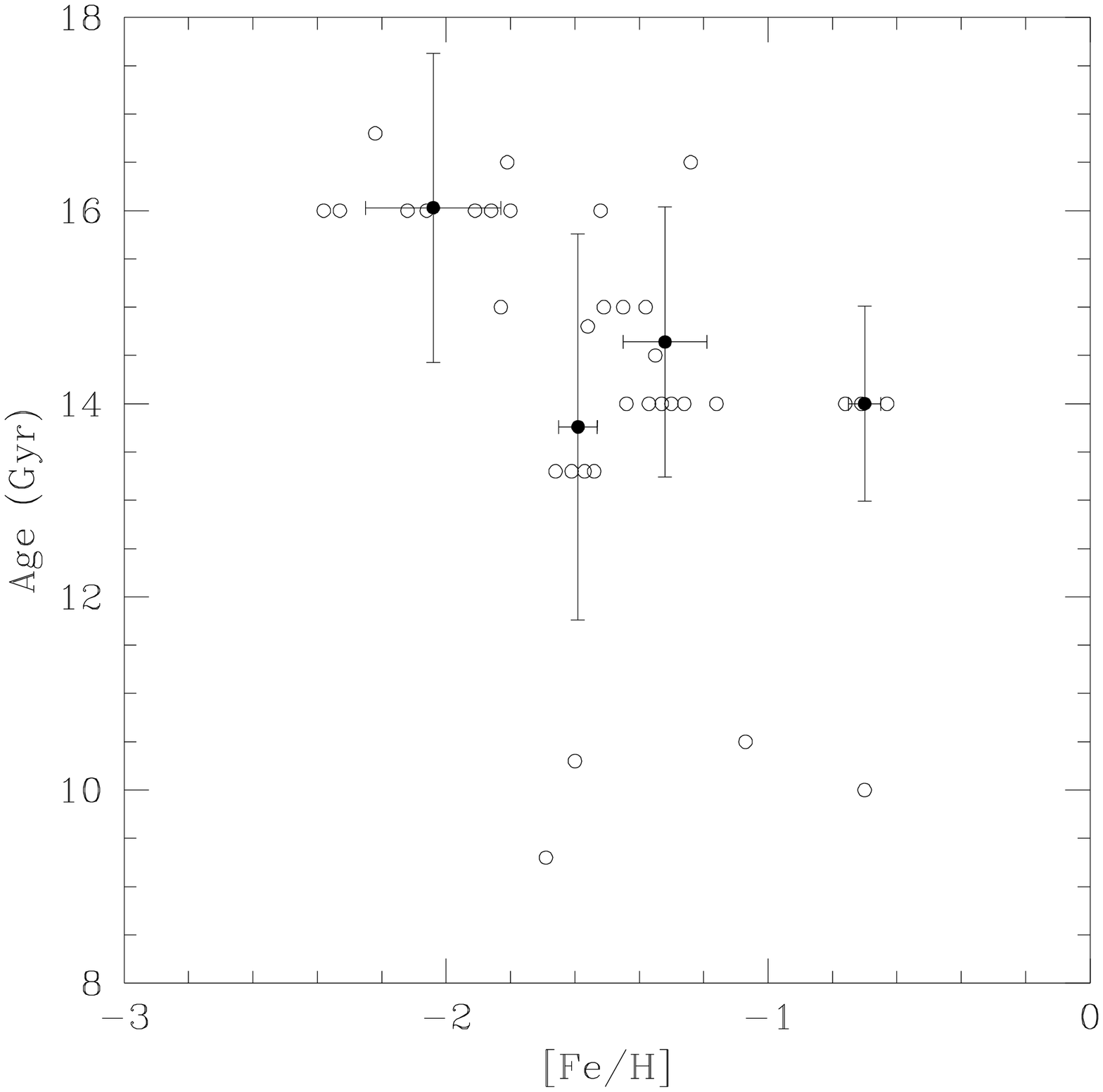}
\figurenum{1}
\caption{The plot of absolute age versus metallicity for the globular clusters
in our sample. Data for the individual clusters are plotted with open
circles with the error bars supressed to avoid cluttering the
diagram. Mean points for each metallicity group with the $\pm
1{\sigma}$ error bars included are plotted as filled circles.}
\end{figure}

\begin{figure}
\plotone{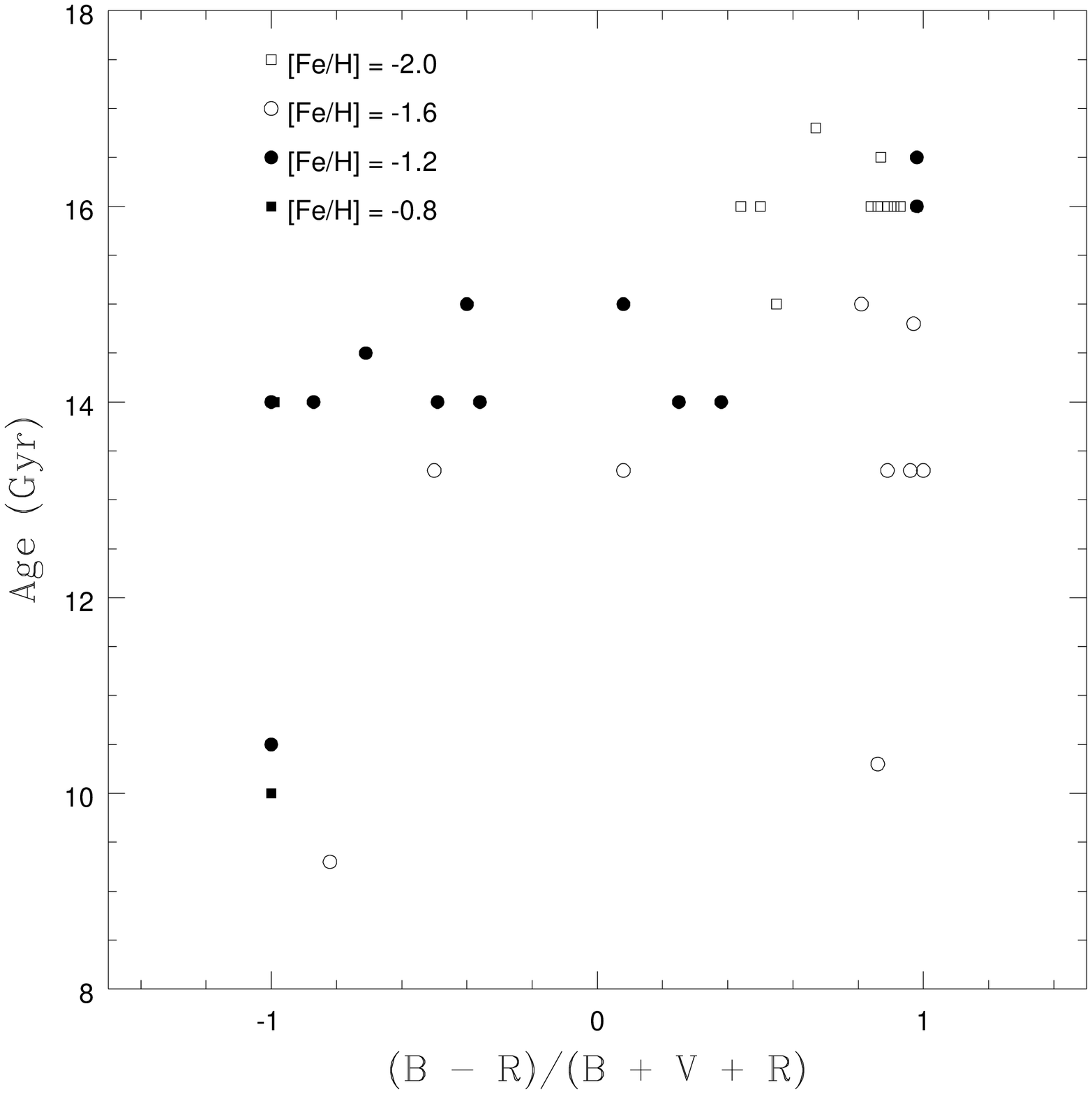}
\figurenum{1}
\caption{The relation between age and HB parameter for the globular clusters.
The symbols correspond to clusters with different metal abundances:
filled squares $[Fe/H] = -0.8$, filled circles $[Fe/H] = -1.2$, open
circles $[Fe/H] = -1.6$, open squares $[Fe/H] = -2.0$.  In the HB
parametrization, pure blue HBs will have an index of $+1$ and pure red
HBs will have an index of $-1$.  Over a broad range in HB types and
metallicities the cluster ages are essentially constant.  This shows
that causes other than metal abundance and age are important in
determining the HB type.}
\end{figure}

\begin{figure}
\plotone{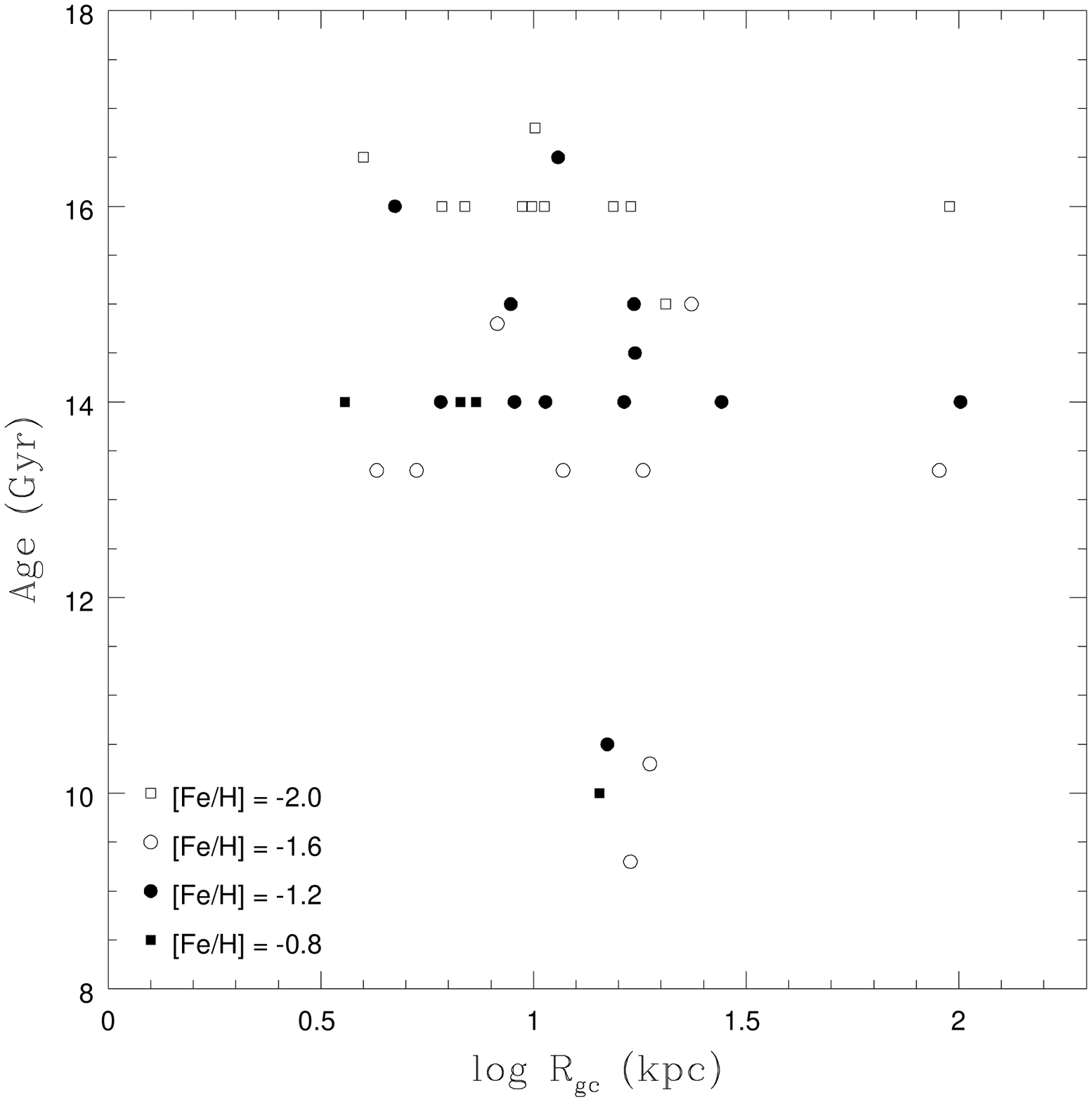}
\figurenum{1}
\caption{Globular cluster ages plotted against Galactocentric
distance.  The different symbols are as in Figure 2. Four clearly
young clusters in the mid-to-outer halo region at ages near 10 Gyr
stand well off the main band of objects. Neglecting these (see text),
there is no evidence of any strong age gradient among the main body of
the cluster system from 4 to 100 kpc from the Galactic center.}
\end{figure}

%%%%%%%%%%%%%%%%%%%%%%%%%%%%%%%%%%%%%%%%%%%%%%%%%%%%%%%%%%%%%%%%%%%%%%
%\renewcommand
%\baselinestretch{1}
\begin{center}
\begin{planotable}{rrrrr}
\tablewidth{4.5in}
\tablecaption{Galactic Globular Cluster Data} \label{tbl-1}
\tablehead
{\colhead{Cluster} &
\colhead{[Fe/H]} &
\colhead{$R_{gc}$(kpc)} &
\colhead{HB-index} &
\colhead{$\Delta Age$(Gyr)}} 
\startdata
%Cluster & [Fe/H] & $R_{gc}$(kpc) & HB-index & $\Delta Age$(Gyr)\nl
  NGC 2298    &-1.86        &15.4    &0.93       &0.0\nl
  NGC 2419    &-2.12        &95.2    &0.86       &0.0\nl
  NGC 4147    &-1.83        &20.5    &0.55       &-1.0\nl
  NGC 4590    &-2.06        &9.9     &0.44       &0.0\nl
  NGC 5053    &-2.38        &16.9    &0.50       &0.0\nl
  NGC 6101    &-1.80        &10.6    &0.84       &0.0\nl
  NGC 6341    &-2.33        &9.4     &0.91       &0.0\nl
  NGC 6397    &-1.91        &6.1     &0.98       &0.0\nl
  NGC 6809    &-1.81        &4.0     &0.87       &+0.5\nl
  NGC 7078    &-2.22        &10.1    &0.67       &+0.8\nl
  NGC 7099    &-2.12        &6.9     &0.89       &0.0\nl
              &             &        &           &   \nl
  NGC 1904    &-1.54        &18.1    &0.89       &-1.7\nl
  NGC 5272    &-1.57        &11.7    &0.08       &-1.7\nl
  NGC 6205    &-1.56        &8.2     &0.97       &-0.2\nl
  NGC 6218    &-1.61        &4.3     &0.96       &-1.7\nl
  NGC 6752    &-1.61        &5.3     &1.00       &-1.7\nl
  NGC 7492    &-1.51        &23.5    &0.81       &0.0\nl
  Rup 106     &-1.69        &16.9    &-0.82      &-5.7\nl
  Arp 2       &-1.6         &18.8    &0.86       &-4.7\nl
  Pal 3       &-1.66        &90.0    &-0.50      &-1.7\nl
              &             &        &           &\nl
  NGC 288     &-1.24        &11.4    &0.98       &+2.5\nl
  NGC 362     &-1.16        &9.0     &-0.87      &0.0\nl
  NGC 1261    &-1.35        &17.3    &-0.71      &+0.5\nl
  NGC 1851    &-1.26        &16.3    &-0.36      &0.0\nl
  NGC 2808    &-1.37        &10.7    &-0.49      &0.0\nl
  NGC 3201    &-1.45        &8.8     &0.08       &+1.0\nl
  NGC 5904    &-1.33        &6.1     &0.38       &0.0\nl
  NGC 6229    &-1.44        &27.7    &0.25       &0.0\nl
  NGC 6254    &-1.52        &4.7     &0.98       &2.0\nl
  Pal 4       &-1.3         &100.9   &-1.00      &0.0\nl
  Pal 5       &-1.38        &17.2    &-0.40      &1.0\nl
  Pal 12      &-1.07        &14.9    &-1.00      &-3.5\nl
              &             &        &           &\nl
  NGC 104     &-0.76        &7.3     &-0.99       &0.0\nl
  NGC 6352    &-0.63        &3.6     &-1.00       &0.0\nl
  NGC 6838    &-0.71        &6.7     &-1.00       &0.0\nl
  Ter 7       &-0.7         &14.3    &-1.00       &-4.0\nl

\tablecomments{Absolute cluster ages can be obtained by adding the fiducial cluster age to each $\Delta Age$. The fiducial cluster for the metal-poorest group is NGC 6341 with an age of 16.0 Gyr. For the next group it is NGC 7492 with an age of 15.0 Gyr. For the group with $[Fe/H] = -1.2$ the standard cluster is NGC 362 at an age of 14.0 Gyr while for the most metal-rich clusters it is NGC 104 at 14.0 Gyr.}
\end{planotable}
\end{center} 

%%%%%%%%%%%%%%%%%%%%%%%%%%%%%%%%%%%%%%%%%%%%%%%%%%%%%%%%%%%%%%%%%%%%%%%%%%%%%

\end{document}